\documentclass[]{spie}

\usepackage{amsmath,amsfonts,amssymb}
\usepackage{graphicx}
\usepackage[colorlinks=true, allcolors=blue]{hyperref}

\title{Population-inversion map of the mesospheric sodium ladder:\\
continuous-wave and pulsed pumping schemes for directed emission}

\author[a,b,*]{Yucheng Yang}
\author[c]{Chunyang Lei}
\author[d]{Kai Guo}
\author[a]{Chi Peng}
\author[a]{Zongpeng Pan}
\affil[a]{State Key Laboratory of Spatial Datum, Beijing 100020, China}
\affil[b]{Beijing Key Laboratory of Quantum Sensing and Precision Measurement, and Center for Quantum Information Technology, and Institute of Quantum Electronics, Peking University, Beijing 100871, China}
\affil[c]{AthenaEyes Co., Ltd., Changsha, China}
\affil[d]{Institute of Systems Engineering, AMS, Beijing 100141, China}

\authorinfo{Correspondence: yangyucheng@pku.edu.cn}

\begin{document}
\maketitle

\begin{abstract}
Directed mirrorless lasing from the mesospheric sodium layer has been proposed
as a way to overcome the isotropy of laser-guide-star fluorescence, with
demonstrated cell-scale analogues and a demonstrated stand-off magnetometry
application. Several transition paths on the Na ladder compete for the same
pump photons, and which of them can sustain a population inversion---and under
what pumping format---has not been classified systematically. We build a
ten-level rate-equation model of the $3S$--$3P$--$4S$--$3D$--$4P$--$4D$ ladder
from NIST transition probabilities and evaluate every electric-dipole line
under four continuous-wave pumping schemes, both in a Doppler-averaged
treatment and in a velocity-selective treatment appropriate for the
collision-poor mesosphere. Three design-relevant results emerge. (1) Under
continuous-wave pumping only three lines can be inverted: $4P_{3/2}\to
4S_{1/2}$ (2.21~\textmu m) and $4P_{3/2}\to 3D_{5/2}$ (9.1~\textmu m), fed
either by direct 330~nm pumping or by the 589+569~nm ladder, and the
fine-structure companion $4S_{1/2}\to 3P_{1/2}$ (1138~nm) under 589+1140~nm
pumping; the frequently discussed $4D_{5/2}\to 4P_{3/2}$ line (2.34~\textmu m)
can never be inverted in steady state because its lower level outlives its
upper level. (2) Velocity selectivity reverses the naive scheme ranking: the
two-step 589+569~nm scheme, which addresses one velocity class at
natural-width cross sections, overtakes direct 330~nm pumping above a total
irradiance of $\sim$40~W\,m$^{-2}$---inside the practically accessible range.
(3) A square-pulse 589+569~nm pump opens a transient inversion window on
2.34~\textmu m that closes within $\sim$200--300~ns, setting a hard upper
bound on useful pulse durations. At practically accessible continuous-wave
irradiances the column-gain exponents remain far below unity, consistent with
published feasibility estimates; the value of the classification is to
identify which lines, schemes, and pulse formats merit further study.
\end{abstract}


\section{INTRODUCTION}
\label{sec:intro}

Sodium laser guide stars (LGS) excite the mesospheric sodium layer at
$\sim$90~km altitude and collect the resonance fluorescence, either for
adaptive optics or, more recently, for remote magnetometry of the geomagnetic
field~\cite{Holzlohner2010,PedrerosBustos2018}. The fluorescence is
spontaneous and therefore nearly isotropic: only a tiny solid-angle fraction
returns to the receiver. If a population inversion could be sustained on a
downward transition of the sodium ladder, amplified spontaneous emission (ASE)
would grow preferentially along the pumped column---a ``mirrorless
laser''---and the emission would become directional. Backward mirrorless
lasing has been demonstrated in air~\cite{Dogariu2011,Hemmer2011}, and in
sodium specifically the cell-scale ingredients exist: pulsed cooperative
backward emission~\cite{Thompson2014}, continuous-wave (CW) mirrorless lasing
at 2.21~\textmu m under 589+569~nm pumping~\cite{Akulshin2018}, correlated
infrared--ultraviolet emission spiking~\cite{Akulshin2021}, and stand-off
magnetometry that reads the geomagnetic field from the directional
emission~\cite{Zhang2021}. A recent review collects the state of the
field~\cite{Akulshin2025}.

\begin{figure}[tb]
\centering
\includegraphics[width=0.72\textwidth]{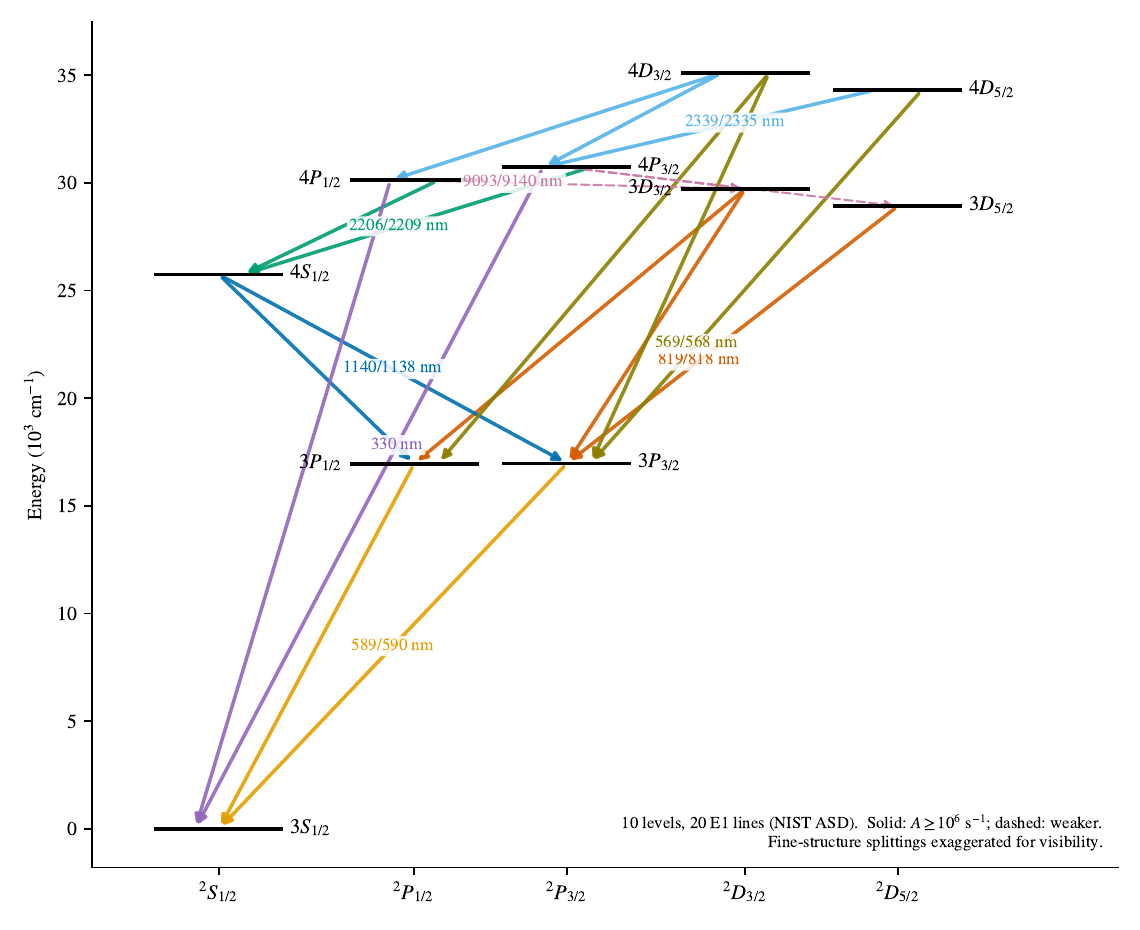}
\caption{The ten fine-structure levels and twenty electric-dipole lines of the
model, drawn from the NIST Atomic Spectra Database~\cite{NISTASD,
Sansonetti2008}. Solid arrows: $A\ge10^{6}$~s$^{-1}$; dashed: weaker lines.
Fine-structure splittings are exaggerated for visibility. Double wavelength
labels (589/590, 569/568, \dots) denote the fine-structure doublets of a
multiplet (e.g., 588.995/589.592~nm for $3P_{3/2}$/$3P_{1/2}\to3S_{1/2}$);
the text refers to each pump scheme by the principal component driven
(589~nm $=$ D2, 569~nm $= 3P_{3/2}\to4D_{5/2}$).}
\label{fig:levels}
\end{figure}

For the mesosphere itself, published feasibility estimates are
sobering. Scaling the cell experiment to the sodium layer puts the ASE
threshold column density an order of magnitude above the natural
$4\times10^{9}$~cm$^{-2}$~\cite{Akulshin2025}, and a survey of mesospheric
metal species found no transition with optical thickness significantly above
one~\cite{Yang2021,Hickson2021}. Those estimates, however, evaluate
\emph{one} scheme at a time. The sodium ladder offers several candidate
paths---the cascade
$4D\to4P\to\{4S,3S,3D\}\to\cdots$ spans lines at 0.33, 0.57, 0.82, 1.14, 2.21,
2.34, and 9.1~\textmu m (Fig.~\ref{fig:levels})---which compete for the same
pump photons, feed each other by spontaneous cascade, and respond differently
to pumping format. Which lines can carry an inversion at all, under which
pumping scheme, CW or pulsed, has to our knowledge not been classified
systematically. That classification is the subject of this paper. We
deliberately keep the model minimal---rate equations with NIST atomic
data---so that every conclusion is traceable to level lifetimes, branching
ratios, and cross sections rather than to model tuning.

\section{MODEL}
\label{sec:model}

\subsection{Levels, lines, and rate equations}

The model comprises the ten fine-structure levels of the
$3S_{1/2}$--$3P_{1/2,3/2}$--$4S_{1/2}$--$3D_{3/2,5/2}$--$4P_{1/2,3/2}$--$4D_{3/2,5/2}$
ladder and the twenty electric-dipole (E1) lines that connect them with
transition probabilities $A\ge10^{3}$~s$^{-1}$, taken from the NIST Atomic
Spectra Database~\cite{NISTASD} (which builds on the critical compilation of
Ref.~\citenum{Sansonetti2008}); transition frequencies are computed from the
NIST level energies. The resulting radiative lifetimes are 16.2/16.3~ns
($3P_{3/2}$/$3P_{1/2}$), 19.4--19.5~ns ($3D$), 37.9~ns ($4S$), 52.3~ns ($4D$),
and 104.7/105.2~ns ($4P_{3/2}$/$4P_{1/2}$)---the $4P$ levels are the
longest-lived rungs of the ladder, which drives much of what follows. The
$4P_{3/2}$ level decays 69.5\% to $4S$ (2206~nm), 28.8\% to $3S$ (330~nm), and
1.7\% to $3D$ (9.1~\textmu m); $4D_{5/2}$ decays 63.3\% to $3P_{3/2}$ (569~nm)
and 36.7\% to $4P_{3/2}$ (2339~nm). Both fine-structure members of the $4P$
and $4D$ terms are retained: the 330.24/330.30~nm pump components are separated
by 168~GHz and select a single member, and one of the inversions found below
lives on an unpumped fine-structure companion line.

Populations $n_i$ (normalized to one atom) obey linear rate equations with
spontaneous decay on all twenty lines and stimulated absorption and emission
on the pumped lines. For a pump of irradiance $I$ on line $l\to u$ the
absorption rate per atom in $l$ is $\sigma_{\rm abs}\,I/h\nu$ with
$\sigma_{\rm abs}=(g_u/g_l)\,\sigma_{\rm se}$, and the model is evaluated in
two ways. In the \emph{Doppler-averaged} treatment, $\sigma_{\rm se} =
(\lambda^{2}/8\pi)\,A\,\phi(0)$ with $\phi(0)$ the peak of the
Doppler-broadened lineshape at $T=185$~K, applied to the whole population. In
the \emph{velocity-selective} treatment (per-velocity-class, in the sense
familiar from velocity-selective saturation spectroscopy), each velocity
class evolves independently---mesospheric velocity-changing collisions occur
on a $\sim$35~\textmu s timescale~\cite{Holzlohner2010}, several hundred
times the longest radiative time, and spontaneous decay preserves the atomic
velocity---with natural-width Lorentzian pump cross sections (peak
$\sigma_0=(\lambda^{2}/2\pi)(g_u/g_l)(A_{ul}/\Gamma_u)$), co-propagating beams
addressing the same class, and the steady state integrated over the Maxwell
distribution. The velocity integral uses a composite grid: a uniform core
spanning $\pm50$ saturation-broadened natural half-widths of the widest pump
line (400 nodes; spacing 0.068~m\,s$^{-1}$ for the 330~nm scheme at
100~W\,m$^{-2}$, resolving the narrowest saturated hole of
Sec.~\ref{sec:ranking} with several nodes per half-width), log-spaced wings
out to $\pm6$ thermal standard deviations (120 nodes per side), and
trapezoidal weights. Doubling both node counts changes the computed
inversions by less than $10^{-4}$ (relative). The two treatments agree to
0.1\% in the linear regime, as they must; they differ, instructively, once
saturation sets in (Sec.~\ref{sec:ranking}).

The figure of merit for a candidate line $u\to l$ is the single-pass
column-gain exponent
\begin{equation}
G \;=\; \sigma_{\rm se}\left(n_u - \tfrac{g_u}{g_l}\,n_l\right) N_{\rm col},
\label{eq:gain}
\end{equation}
evaluated at the natural sodium column density $N_{\rm col} =
4\times10^{13}$~m$^{-2}$~\cite{Akulshin2025}. $G$ is the column optical
thickness of the inverted line at line center: in the notation of
Ref.~\citenum{Akulshin2025} (Sec.~5.4), which defines a net emission length
$\ell_i=\Gamma_D/(n_i\sigma\Gamma_2)$ averaged over the spontaneous-emission
spectrum, $G$ equals $L/\ell_i$ up to the line-center/line-average
convention, and the two coincide as threshold criteria in the
spontaneous-emission-dominated limit.

\subsection{Assumptions and validity}
\label{sec:assumptions}

The model is deliberately minimal. Hyperfine and Zeeman structure are not
resolved; collisional quenching, spin randomization, and photon recoil are
neglected; the pump is monochromatic and resonant; and radiative transport
(ASE saturation) is not treated---Eq.~(\ref{eq:gain}) is a small-signal
exponent, not a flux prediction. Two consequences must be stated
plainly. First, all irradiances quoted below are \emph{model} parameters: for
CW illumination of the mesosphere, practically achievable irradiances top out
at roughly $10^{2}$~W\,m$^{-2}$ (a 50-W-class launcher focused to a
$\sim$1-m-diameter spot); results at higher CW irradiance are model
exploration only, shown to expose the physics, not to suggest reachable
operating points. Pulsed lasers, by contrast, reach far higher \emph{peak}
irradiance at low duty cycle, which is what Sec.~\ref{sec:pulsed}
exploits. Second, the velocity-selective treatment brackets reality from the
collisionless side; velocity-changing collisions would slowly refill the
pumped class and push results toward the Doppler-averaged limit. The
$\sim$35~\textmu s collision timescale of Ref.~\citenum{Holzlohner2010} is
estimated for ground-state atoms on the D2 line; the high-lying working
levels are geometrically larger, and scaling the collision cross section as
$n^{*4}$ (effective quantum numbers $n^{*}=2.12$, 3.13, and 3.99 for
$3P_{3/2}$, $4P_{3/2}$, and $4D_{5/2}$) multiplies it by $\sim$5 for $4P$
and $\sim$13 for $4D$. Even so, the class-refill and collision times remain
$\gtrsim$3~\textmu s---more than an order of magnitude above the longest
radiative lifetime (105~ns)---so the collisionless steady states used here
are not materially perturbed; measured quenching and velocity-changing cross
sections for these Rydberg-like levels under mesospheric N$_2$/O$_2$
conditions are, to our knowledge, not available, and Ref.~\citenum{Akulshin2025}
(Sec.~5.5) lists the same collision channels as open modeling questions for
any sky experiment.

Internal validation is built into every run: the computed lifetimes and
branching ratios reproduce the NIST values; with the pump off, all population
returns to $3S_{1/2}$; a saturating D2 pump drives $n_{3P_{3/2}}/n_{3S}$ to
the degeneracy ratio of 2; and the velocity-integrated linear-regime
excitation matches the Doppler-averaged model to 0.1\%.

\section{CONTINUOUS-WAVE CLASSIFICATION}
\label{sec:cw}

\subsection{Which lines can be inverted at all}

Four CW schemes cover the pump-accessible paths: S1, direct 330~nm pumping
$3S\to4P_{3/2}$; S2, two-step $589+569$~nm pumping $3S\to3P_{3/2}\to4D_{5/2}$
(the scheme of the cell demonstration~\cite{Akulshin2018}); S3, two-step
$589+820$~nm pumping $3S\to3P_{3/2}\to3D_{5/2}$; and S4, two-step
$589+1140$~nm pumping $3S\to3P_{3/2}\to4S_{1/2}$. In the two-step schemes the
total irradiance is split equally between the beams; a scan of the split
fraction puts the optimum at 0.40--0.55 of the total in the second step
throughout the practically accessible range, so the equal split loses almost
nothing. Table~\ref{tab:classification} summarizes which lines carry a
positive steady-state inversion.

\begin{table}[tb]
\caption{CW classification. ``Inverted lines'' are those with a positive
steady-state inversion under the scheme; $G$ is the velocity-selective
column-gain exponent, Eq.~(\ref{eq:gain}), of the strongest line at a total
irradiance of 100~W\,m$^{-2}$.}
\label{tab:classification}
\centering
\begin{tabular}{llll}
\hline
Scheme & Pump path & Inverted lines & $G$ at 100~W\,m$^{-2}$ \\
\hline
S1: 330~nm & $3S\to4P_{3/2}$ &
  $4P_{3/2}\to3D_{5/2,3/2}$ (9.1~\textmu m), & $1.6\times10^{-5}$ \\
& & $4P_{3/2}\to4S$ (2206~nm), $4S\to3P_{3/2,1/2}$ & \\
S2: 589+569~nm & $3S\to3P_{3/2}\to4D_{5/2}$ &
  same $4P_{3/2}$ lines, via cascade & $3.0\times10^{-5}$ \\
S3: 589+820~nm & $3S\to3P_{3/2}\to3D_{5/2}$ & none & --- \\
S4: 589+1140~nm & $3S\to3P_{3/2}\to4S$ &
  $4S\to3P_{1/2}$ (1138~nm) & $1.9\times10^{-5}$ \\
\hline
\multicolumn{4}{l}{$4D_{5/2}\to4P_{3/2}$ (2339~nm): \emph{never inverted in
steady state, under any scheme or irradiance.}} \\
\hline
\end{tabular}
\end{table}

Three structural facts, each traceable to the atomic data, explain the
table. First, the $4P$ levels are the natural CW inversion reservoir: at
$\sim$105~ns they outlive their lower levels $4S$ (37.9~ns) and $3D$
(19.5~ns), so populating $4P$ by any route inverts both the 2206~nm and the
9.1~\textmu m lines---direct 330~nm pumping and the $4D\to4P$ cascade of the
589+569~nm scheme both work, consistent with the CW cell
demonstration~\cite{Akulshin2018}. Second, and by the same token, the
2339~nm line can \emph{never} be CW-inverted: its lower level $4P_{3/2}$
(105~ns) outlives its upper level $4D_{5/2}$ (52~ns) and receives 37\% of the
upper level's decay, so the degeneracy-weighted inversion is negative at
every irradiance---we verified this up to $10^{7}$~W\,m$^{-2}$. Third, S4
inverts the \emph{unpumped} fine-structure companion: pumping $3P_{3/2}\to4S$
clamps that pair, but $4S\to3P_{1/2}$ (1138~nm) sees a lower level that the
D2 pump continuously empties into the ground state---an inversion pattern
familiar from polychromatic mirrorless lasing in cesium
vapor~\cite{Antypas2019}. S3, the one scheme that inverts nothing, makes the
lifetime logic quantitative by counterexample: its terminal level $3D_{5/2}$
(19.4~ns) outlives the lower level $3P_{3/2}$ (16.2~ns) by a factor of only
1.20---compare 2.76 for the $4P/4S$ pair that carries the inversions
above---and the degeneracy-weighted condition $n_u>(g_u/g_l)\,n_l$ must be
met with $g_u/g_l=6/4$, so the marginal lifetime edge is not enough.

\subsection{Velocity selectivity reverses the scheme ranking}
\label{sec:ranking}

\begin{figure}[tb]
\centering
\begin{minipage}{0.49\textwidth}
\centering
\includegraphics[width=\textwidth]{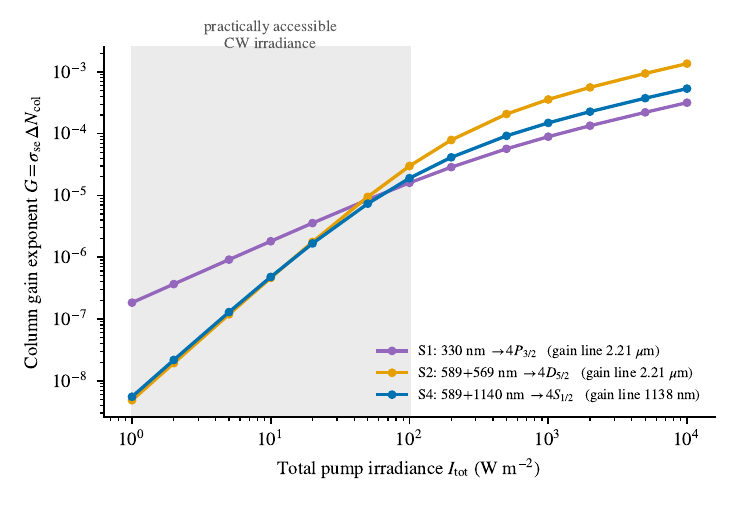}
\end{minipage}
\begin{minipage}{0.49\textwidth}
\centering
\includegraphics[width=\textwidth]{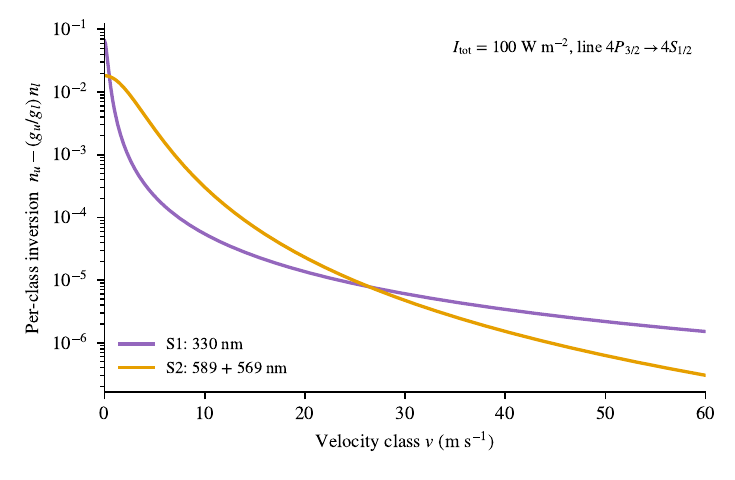}
\end{minipage}
\caption{Left: velocity-selective column-gain exponent of the strongest line
of each inverting scheme versus total pump irradiance. The shaded band marks
the practically accessible CW range; the S1/S2 crossover falls inside it, at
$\sim$40~W\,m$^{-2}$. Right: the mechanism---per-velocity-class inversion of
the 2206~nm line at 100~W\,m$^{-2}$ total. The 330~nm scheme saturates a
$\sim$0.3~m\,s$^{-1}$-wide class; the 589+569~nm scheme burns a hole an order
of magnitude wider, because the D2 first step has an 11-fold larger natural
cross section and a 3-fold lower class-saturation irradiance.}
\label{fig:ranking}
\end{figure}

In the Doppler-averaged treatment, direct 330~nm pumping wins at every
irradiance: one photon per excitation, no intermediate step. The
velocity-selective treatment (Fig.~\ref{fig:ranking}) tells a different and
more interesting story. At low irradiance S1 still leads (at
10~W\,m$^{-2}$: $G=1.8\times10^{-6}$ versus $4.6\times10^{-7}$ for S2),
because the two-step excitation is quadratic in the pump. But the two-step
scheme overtakes at $\sim$40~W\,m$^{-2}$ total, and at 100~W\,m$^{-2}$ it
leads $3.0\times10^{-5}$ to $1.6\times10^{-5}$---the crossover sits squarely
inside the practically accessible decade.

The mechanism is hole burning. In the collision-poor mesosphere both pump
beams talk to the \emph{same} velocity class, so the second step acts, at the
natural-width cross section, on exactly the population that the first step
concentrated---the two-step scheme pays no double Doppler penalty. What
differs between schemes is the \emph{width} of the saturated hole: the D2
first step has $\sigma_0=1.1\times10^{-13}$~m$^{2}$ against
$1.0\times10^{-14}$~m$^{2}$ for the 330~nm line, and a per-class saturation
irradiance of 188 against 574~W\,m$^{-2}$. At 100~W\,m$^{-2}$ the 330~nm
scheme has saturated a class only $\sim$0.3~m\,s$^{-1}$ wide (per-class
inversion 0.065 at line center), while the 589+569~nm scheme sustains a
per-class inversion of 0.018 across a $\sim$2.5~m\,s$^{-1}$ hole
(Fig.~\ref{fig:ranking}, right); the wider hole wins the velocity integral.
A Doppler-averaged estimate misses this entirely---it underestimates the
two-step scheme by up to two orders of magnitude in the linear regime.

\section{PULSED PUMPING: THE 2339~nm WINDOW}
\label{sec:pulsed}

\begin{figure}[tb]
\centering
\includegraphics[width=0.6\textwidth]{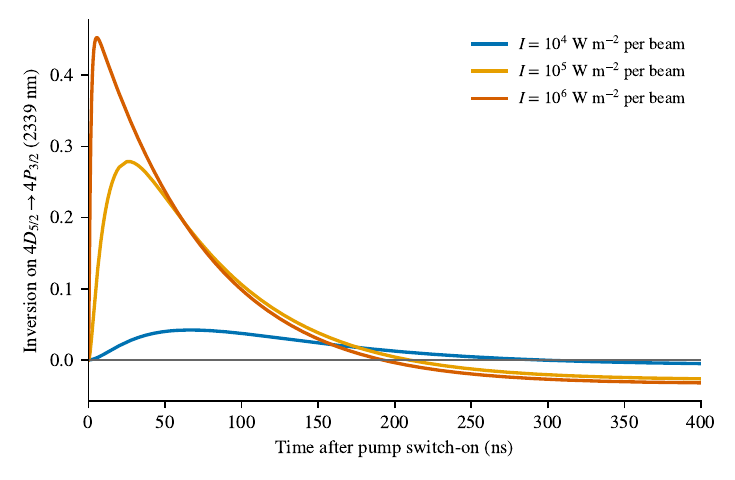}
\caption{Transient inversion of the CW-forbidden $4D_{5/2}\to4P_{3/2}$
(2339~nm) line after a square 589+569~nm pump switches on at $t=0$ (equal
irradiance per beam; Doppler-peak treatment). The window closes within
$\sim$200--300~ns at all pump levels as the long-lived $4P_{3/2}$ reservoir
fills.}
\label{fig:pulsed}
\end{figure}

The steady-state prohibition of the 2339~nm line is a statement about
reservoir levels, not about the line itself, and it therefore has a loophole:
\emph{before} the $4P_{3/2}$ reservoir fills, the line is invertible. A
square 589+569~nm pump switched on at $t=0$ fills $4D_{5/2}$ within
nanoseconds at high irradiance, while $4P_{3/2}$ needs on the order of its
105~ns lifetime to accumulate. The transient calculation integrates the full
ten-level system from the ground state, so the sequential buildup of the
intermediate $3P_{3/2}$ population is part of the solution rather than an
assumption: at $10^{4}$~W\,m$^{-2}$ per beam---160 times the D2 saturation
irradiance of 62.6~W\,m$^{-2}$~\cite{Steck2010}---$3P_{3/2}$ reaches 90\% of
its quasi-steady value within 14~ns (and proportionally faster at higher
irradiance), well before the inversion peaks. Figure~\ref{fig:pulsed}
quantifies the resulting window: the peak transient inversion per atom
reaches 0.042 at $10^{4}$~W\,m$^{-2}$ per beam (peaking at 67~ns), 0.28 at
$10^{5}$ (26~ns), and 0.45 at $10^{6}$ (5.4~ns), and the inversion turns
negative by $\sim$290, 210, and 190~ns respectively. The design rule is a
hard one: any attempt to exploit the 2339~nm path must use pulses shorter
than $\sim$200~ns---and captures most of the available inversion only below
$\sim$100~ns---with a leading edge short compared to the $\sim$14~ns
first-step buildup, since a slower turn-on delays and shallows the peak the
square-pulse idealization predicts. The required peak irradiances are
naturally provided by pulsed lasers at low duty cycle. Transient-inversion physics of this kind is
experimentally established in sodium cells: pulsed excitation produces
backward cooperative emission at 1140/1138~nm~\cite{Thompson2014}, whereas
the CW demonstration occurred on the reservoir-fed 2.21~\textmu m
line~\cite{Akulshin2018}; the 2339~nm window predicted here awaits a
dedicated test.

\section{DISCUSSION}
\label{sec:discussion}

\textbf{Feasibility in context.} At practically accessible CW irradiances the
largest column-gain exponent found here is $G\sim3\times10^{-5}$
(Table~\ref{tab:classification}): single-pass amplification of
$e^{G}-1\approx3\times10^{-5}$, i.e., no mesospheric mirrorless laser under
any CW scheme at the natural column density---in independent agreement with
the published feasibility verdicts~\cite{Akulshin2025,Yang2021,Hickson2021}.
The contribution of this classification is therefore not a feasibility claim
but a map: it identifies the two lines worth watching (2206~nm and
9.1~\textmu m, both fed by the $4P$ reservoir), the scheme that uses
practically available power most efficiently (589+569~nm above
$\sim$40~W\,m$^{-2}$---conveniently, the scheme with the most mature laser
technology, since both wavelengths are available from
sodium-guide-star-adjacent sources), the one line that requires pulsed
operation (2339~nm, pulses $\lesssim$200~ns), and the schemes not worth
pursuing (589+820~nm for inversion purposes; 330~nm alone at high power,
where its narrow burned hole wastes pump photons).

\textbf{Contact with the cell experiment.} The one laboratory anchor for
this classification is the CW cell demonstration of
Ref.~\citenum{Akulshin2018}, and the comparison is quantitative, not merely
consistent. Applied to the reported threshold conditions of that
experiment---summarized against the mesospheric parameters in
Table~\ref{tab:cell}---the same velocity-selective model yields $G\approx3$
on the 2206~nm line: it locates the observed threshold to within a factor of
three of $G=1$, which is as close as a model without hyperfine or Zeeman
structure should be expected to come (the experiment relies on a 1.7-GHz
repumping sideband, and the richer density-matrix estimate of
Ref.~\citenum{Akulshin2025}, Sec.~5.4, places the threshold column an order
of magnitude below the measured one). The comparison also dissolves an
apparent contradiction: lasing at ``sub-10~mW total power'' is not lasing at
low irradiance---focused to 250--350~\textmu m spots, the threshold-scan
powers of 30 and 12~mW correspond to $6\times10^{5}$ and
$1.2\times10^{5}$~W\,m$^{-2}$, three to four orders of magnitude above the
practically accessible mesospheric CW irradiance, applied to a column
800 times denser than the sodium layer. The cell values of
Table~\ref{tab:classification} and the cell demonstration therefore occupy
the same model surface at very different coordinates. One caution transfers
in the opposite direction: Ref.~\citenum{Akulshin2018} also reports weak
directional light at 2.34~\textmu m under CW pumping, roughly ten times
weaker than the 2.21~\textmu m emission. Our steady-state prohibition
concerns population inversion only; parametric four-wave mixing, which
requires no inversion and lies outside any rate-equation description, is the
natural reading of that observation, consistent with the FWM--ASE interplay
those authors discuss.

\begin{table}[tb]
\caption{The vapor-cell threshold conditions of Ref.~\citenum{Akulshin2018}
against the mesospheric parameters used in this paper. Cell irradiances are
the quoted powers over the quoted focal-spot areas; the cell class-refill
time is the transverse beam-transit time, the mesospheric one the
velocity-changing collision time~\cite{Holzlohner2010}.}
\label{tab:cell}
\centering
\begin{tabular}{lll}
\hline
Parameter & Vapor cell~\cite{Akulshin2018} & Mesosphere (this work) \\
\hline
Temperature & 458~K & 185~K \\
Na column density & $3.2\times10^{12}$~cm$^{-2}$ ($3.2\times10^{11}$~cm$^{-3}\times10$~cm) & $4\times10^{9}$~cm$^{-2}$ \\
Pump irradiance (589/569) & $6\times10^{5}$ / $1.2\times10^{5}$~W\,m$^{-2}$ & $\lesssim10^{2}$~W\,m$^{-2}$ total (CW) \\
Gain length / diameter & 0.1~m / 250--350~\textmu m & $\sim$10$^{4}$~m / $\sim$1~m \\
Aspect ratio & $2.5\times10^{-3}$ & $\sim$10$^{-4}$ \\
Class-refill time & $\sim$0.4~\textmu s (transit) & $\sim$35~\textmu s (collisions) \\
Model $G$ (2206~nm) & $\approx$3 at threshold & $3\times10^{-5}$ at $10^{2}$~W\,m$^{-2}$ \\
\hline
\end{tabular}
\end{table}

\textbf{Limitations.} The caveats of Sec.~\ref{sec:assumptions} bound all
quantitative statements. Hyperfine structure redistributes oscillator
strength within each line and will modify saturation behavior---and it can
do more than redistribute: ground-state hyperfine pumping opens dark states,
and in S4 the $F=1$ component of the D2 step feeds population paths the
fine-structure model does not resolve, so the 1138~nm companion inversion in
particular should be re-examined with hyperfine structure before being
relied on. Collisional quenching and spin randomization remove population
and coherence on microsecond timescales (with the excited-state caveat of
Sec.~\ref{sec:assumptions}); photon recoil slowly reshapes the velocity
distribution under strong CW pumping; and an actual ASE flux prediction
requires radiative transport along the column. The Zeeman response of the
inverted medium---the property that makes directional emission a
magnetometer~\cite{Zhang2021}---is deliberately left to follow-up work, as is
the question of how far frequency-chirped pumping, which compresses the
velocity distribution, can raise the effective participation
fraction~\cite{Akulshin2025}.

\section{CONCLUSIONS}
\label{sec:conclusions}

A minimal NIST-data rate model classifies the mirrorless-lasing candidates of
the mesospheric sodium ladder as follows: CW inversion exists only on the
$4P$-fed lines at 2206~nm and 9.1~\textmu m and on the fine-structure
companion at 1138~nm; the 2339~nm line is CW-forbidden by lifetime ordering
and pulse durations must stay below $\sim$200~ns to use it; velocity
selectivity makes the 589+569~nm ladder the most power-efficient CW scheme
above $\sim$40~W\,m$^{-2}$ total irradiance, reversing the Doppler-averaged
ranking; and all CW column gains at accessible irradiance remain far below
threshold at the natural sodium column density. The model, its NIST-derived
inputs, and every figure of this paper are generated by scripts with built-in
validation against the atomic data. The analytic counterpart of this
survey---closed-form criteria that explain the classification, the scheme
ranking, and the pulse-length limit from the atomic constants alone---is
developed in a separate paper\cite{YangLei2026nantong}.

\acknowledgments
This work was supported by the National Natural Science Foundation of China
under Grant No.~62301377.

\bibliography{refs}

@article{Holzlohner2010,
  author  = {Holzl{\"o}hner, R. and Rochester, S. M. and Bonaccini Calia, D. and Budker, D. and Higbie, J. M. and Hackenberg, W.},
  title   = {Optimization of cw sodium laser guide star efficiency},
  journal = {Astronomy \& Astrophysics},
  volume  = {510},
  pages   = {A20},
  year    = {2010}
}

@article{PedrerosBustos2018,
  author  = {Pedreros Bustos, F. and Bonaccini Calia, D. and Budker, D. and Centrone, M. and Hellemeier, J. and Hickson, P. and Holzl{\"o}hner, R. and Rochester, S.},
  title   = {Remote sensing of geomagnetic fields and atomic collisions in the mesosphere},
  journal = {Nature Communications},
  volume  = {9},
  pages   = {3981},
  year    = {2018}
}

@article{Akulshin2025,
  author  = {Akulshin, A. M. and Budker, D. and Pedreros Bustos, F. and Dang, T. and Klinger, E. and Rochester, S. M. and Wickenbrock, A. and Zhang, R.},
  title   = {Remote detection optical magnetometry},
  journal = {Physics Reports},
  volume  = {1106},
  pages   = {1--32},
  year    = {2025}
}

@article{Dogariu2011,
  author  = {Dogariu, A. and Michael, J. B. and Scully, M. O. and Miles, R. B.},
  title   = {High-gain backward lasing in air},
  journal = {Science},
  volume  = {331},
  pages   = {442--445},
  year    = {2011}
}

@article{Hemmer2011,
  author  = {Hemmer, P. R. and Miles, R. B. and Polynkin, P. and Siebert, T. and Sokolov, A. V. and Sprangle, P. and Scully, M. O.},
  title   = {Standoff spectroscopy via remote generation of a backward-propagating laser beam},
  journal = {Proceedings of the National Academy of Sciences},
  volume  = {108},
  number  = {8},
  pages   = {3130--3134},
  year    = {2011}
}

@article{Thompson2014,
  author  = {Thompson, J. V. and Ballmann, C. W. and Cai, H. and Yi, Z. and Rostovtsev, Y. V. and Sokolov, A. V. and Hemmer, P. and Zheltikov, A. M. and Ariunbold, G. O. and Scully, M. O.},
  title   = {Pulsed cooperative backward emissions from non-degenerate atomic transitions in sodium},
  journal = {New Journal of Physics},
  volume  = {16},
  pages   = {103017},
  year    = {2014}
}

@article{Akulshin2018,
  author  = {Akulshin, A. M. and Pedreros Bustos, F. and Budker, D.},
  title   = {Continuous-wave mirrorless lasing at 2.21~{\textmu}m in sodium vapors},
  journal = {Optics Letters},
  volume  = {43},
  number  = {21},
  pages   = {5279--5282},
  year    = {2018}
}

@article{Akulshin2021,
  author  = {Akulshin, A. and Pedreros Bustos, F. and Budker, D.},
  title   = {Intensity-correlated spiking of infrared and ultraviolet emission from sodium vapors},
  journal = {Optics Letters},
  volume  = {46},
  number  = {9},
  pages   = {2131--2134},
  year    = {2021}
}

@article{Antypas2019,
  author  = {Antypas, D. and Tretiak, O. and Budker, D. and Akulshin, A.},
  title   = {Polychromatic, continuous-wave mirrorless lasing from monochromatic pumping of cesium vapor},
  journal = {Optics Letters},
  volume  = {44},
  number  = {14},
  pages   = {3657--3660},
  year    = {2019}
}

@article{Zhang2021,
  author  = {Zhang, R. and Klinger, E. and Pedreros Bustos, F. and Akulshin, A. and Guo, H. and Wickenbrock, A. and Budker, D.},
  title   = {Stand-off magnetometry with directional emission from sodium vapors},
  journal = {Physical Review Letters},
  volume  = {127},
  pages   = {173605},
  year    = {2021}
}

@article{Yang2021,
  author  = {Yang, R. and Hellemeier, J. and Hickson, P.},
  title   = {Atomic transitions for adaptive optics},
  journal = {Journal of the Optical Society of America B},
  volume  = {38},
  number  = {8},
  pages   = {2239--2251},
  year    = {2021}
}

@article{Hickson2021,
  author  = {Hickson, P. and Hellemeier, J. and Yang, R.},
  title   = {Can amplified spontaneous emission produce intense laser guide stars for adaptive optics?},
  journal = {Optics Letters},
  volume  = {46},
  number  = {8},
  pages   = {1792--1795},
  year    = {2021}
}

@article{Sansonetti2008,
  author  = {Sansonetti, J. E.},
  title   = {Wavelengths, transition probabilities, and energy levels for the spectra of sodium ({Na~I}--{Na~XI})},
  journal = {Journal of Physical and Chemical Reference Data},
  volume  = {37},
  number  = {4},
  pages   = {1659},
  year    = {2008}
}

@misc{NISTASD,
  author       = {{NIST ASD Team}},
  title        = {{NIST} Atomic Spectra Database},
  howpublished = {\url{https://physics.nist.gov/asd}},
  note         = {National Institute of Standards and Technology, Gaithersburg, MD; retrieved 16 July 2026}
}

@misc{Steck2010,
  author       = {Steck, D. A.},
  title        = {Sodium {D} Line Data},
  howpublished = {available online at \url{http://steck.us/alkalidata}},
  note         = {revision 2.1.4, 23 December 2010}
}

@misc{YangLei2026nantong,
  author = {Yang, Y. and Lei, C. and Guo, K. and Peng, C.},
  title  = {Which transition can be used in sodium mirrorless lasing for mesospheric magnetometry?},
  eprint={2608.23272},
  archivePrefix={arXiv},
  primaryClass={physics.app-ph},
  howpublished = {\url{https://arxiv.org/abs/2608.23272}},
  year={2026}
}
\bibliographystyle{spiebib}

\end{document}